\documentclass[11pt]{article}  
 \pdfoutput=1 
 \usepackage{jheppub}
 \usepackage{epsfig}
\usepackage{graphicx, color, xcolor}
\usepackage{amsmath,amsfonts,amssymb}
\usepackage{slashed}

 \usepackage{subcaption,caption}
\usepackage{tcolorbox}

\newcommand{\beq}{\begin{equation}}
\newcommand{\eeq}{\end{equation}}
\newcommand{\bea}{\begin{eqnarray}}
\newcommand{\eea}{\end{eqnarray}}
\newcommand{\be}{\begin{equation}}
\newcommand{\ee}{\end{equation}}

\title{Renormalisation Group approach to pandemics as a time-dependent SIR model}
\author{Michele Della Morte$^{1}$,}\author{Francesco Sannino$^{2,3}$}
\affiliation{$^{1}$ IMADA \& CP3-Origins. Univ. of Southern Denmark, Campusvej 55, DK-5230 Odense, Denmark}
  \affiliation{$^{2} $CP$^3$-Origins and D-IAS, Univ. of Southern Denmark,  Campusvej 55, DK-5230 Odense, Denmark}
  \affiliation{$^{3}$ Dipartimento di Fisica, E. Pancini, Univ. di Napoli, Federico II and INFN sezione di Napoli \\  Complesso Universitario di Monte S. Angelo Edificio 6, via Cintia, 80126 Napoli, Italy }
  \abstract{  We generalise the epidemic Renormalisation Group framework while connecting it to a SIR model with time-dependent coefficients.  
  We then  confront the model  with COVID-19 in Denmark, Germany, Italy and France and show that the   approach works rather well in reproducing the data. We also show that a better understanding of the time dependence of the recovery rate  would require extending the model to take into account the number of deaths whenever these are over 15\% of the total number of infected cases. }

\begin{document}
 \maketitle

\section{Introduction}
%

Epidemic dynamics is often described in terms of a simple model introduced long time ago in \cite{Kermack:1927}. Here, the affected population is described in terms
of compartmentalised sub-populations that have different roles in the dynamics. Then, differential equations are designed to describe the
time evolution of the various compartments. 
The sub-populations
can be chosen to represent (S)usceptible, (I)nfected and (R)ecovered individuals (SIR model), obeying the following differential equations:
\begin{eqnarray}
\label{SESE}
\frac{dS}{dt} &=& -\gamma S \frac{I}{N} \;, \nonumber \\
\frac{dI}{dt} &=& \gamma S \frac{I}{N} - \varepsilon I \;,  \\
\frac{dR}{dt} &=& \varepsilon I\;, \nonumber
\end{eqnarray}
with the conservation law 
\begin{equation}
\label{conservation}
S(t) + I(t) + R(t) = N \ .
\end{equation} 
The system depends on three parameters, namely $\gamma$, $\varepsilon$ and $N$. Due to the conservation law \eqref{conservation}, only two equations are independent, so that one can drop the equation for $S$.

 The total number of infected, $\tilde{I} (t)$, that we are interested in, is related to the above sub-populations as
\begin{equation}
\tilde{I} (t) = I(t) + R(t)\,.
\end{equation}
We can therefore re-write the two independent SIR equations as
\begin{eqnarray}
\frac{d \tilde{I} (t)}{d t} &=& {\gamma} \left(\tilde{I} (t) - R(t) \right) \left( 1 - \frac{\tilde{I}(t)}{N} \right)\,, \label{eq:SIR1}\\
\frac{d R (t)}{d t} &=& \varepsilon \left( \tilde{I} (t) - R(t) \right)\ . \label{eq:SIR2}
\end{eqnarray} 

Empirical modifications of the basic SIR model exist and range from including new sub-populations to generalise the coefficients $\gamma$, $\varepsilon$ to be time-dependent in order to better reproduce the observed data. 

Recently the epidemic Renormalisation Group approach (eRG) to pandemics, inspired by particle physics methodologies,  was put forward in \cite{DellaMorte:2020wlc} and further explored in \cite{Cacciapaglia:2020mjf}. In the latter paper it was demonstrated the eRG effectiveness when describing how the pandemic spreads across different regions of the world. 

The goal of the present work is to further extend the eRG framework  to properly take into account the number of recovered cases so that a better   understanding of the reproduction number can also be achieved. We will start, first, by providing a map between the original eRG model and certain modified SIR models. We will finally test the framework via COVID-19 data.   

\subsection{Reviewing the eRG}
In the \emph{epidemic renormalisation group (eRG)} approach \cite{DellaMorte:2020wlc}, rather than the number of cases, it is convenient to discuss its logarithm, which is a more slowly varying function  \begin{equation} 
\alpha(t) = \ln\ \tilde{I}(t) \ ,
\end{equation}  where $\ln$ indicates the natural logarithm.  The derivative of $\alpha$ with respect to time provides a new quantity that we interpret as the {\it beta-function} of an underlying microscopic model. In statistical and high energy physics, the latter governs the time (inverse energy) dependence of  the interaction strength among fundamental particles. Here it regulates infectious interactions.

 More specifically, as the renormalisation group equations in high energy physics are expressed in terms of derivatives with respect to the energy $\mu$, it is natural to identify the time as
  $t/t_0=-\ln ({\mu/\mu_0})$, where $t_0$ and $\mu_0$ are respectively a reference time and energy scale. We choose $t_0$ to be one week so that time is measured in weeks, and will drop it in the following. 
 Thus, the dictionary between the eRG equation for the epidemic strength $\alpha$ and the high-energy physics analog is
 \begin{equation}
 \label{betaalpha}
 \beta (\alpha) = \frac{d \alpha}{d\ln\left(\mu/\mu_0\right)} = - \frac{d \alpha}{dt} \ . 
\end{equation} 
 It has been shown in \cite{DellaMorte:2020wlc} that $\alpha$ captures the essential information about the infected population within a sufficiently isolated region of the world.
The pandemic beta function can be parametrised as
\beq
- \beta (\alpha) = \frac{d \alpha}{dt} = \tilde{\gamma} \, \alpha   \left( 1 - \frac{\alpha}{a} \right)^n\,,
\label{eq:beta0}
\eeq 
 whose solution, for $n=1$, is a familiar logistic-like function
\begin{equation}
 \alpha (t) =  \frac{a e^{\tilde{\gamma} t}}{b + e^{\tilde{\gamma} t}}\,.
\end{equation}
The dynamics encoded in Eq.~\eqref{eq:beta0} is that of a system that flows from an UV fixed point at $t=-\infty$ where $\alpha = 0$  to an IR fixed point where $\alpha = a$. The latter value encodes the total number of infected cases $P = \exp (a)$ in the region under study. The coefficient $\tilde{\gamma}$ is the diffusion slope, while $b$ shifts the entire epidemic curve by a given amount of time. Further details, including what parameter influences the {\it flattening of the curve} and location of the inflection point and its properties can be found in \cite{DellaMorte:2020wlc}. 

\subsection{Connecting eRG with SIR while extending it}
To start connecting with compartmental models we rewrite  Eq.~\eqref{betaalpha} as

\be
\frac{d\tilde{I}}{dt}=\tilde{\gamma} \tilde{I} \ln\tilde{I} \left( 1- \frac{\ln\tilde{I}}{\ln P} \right)\;,
\ee
whose solution, with the initial condition $\ln \tilde{I}_0=\ln \tilde{I}(0)=\frac{\ln P}{b+1}$, is a logistic function written as
\be
\tilde{I}(t)=\exp\left({a \frac{1}{b e^{-\tilde{\gamma}t}+1}}\right) \;.
\ee

\begin{figure}[h!]
    \begin{center}
    \includegraphics[width=0.7\textwidth]{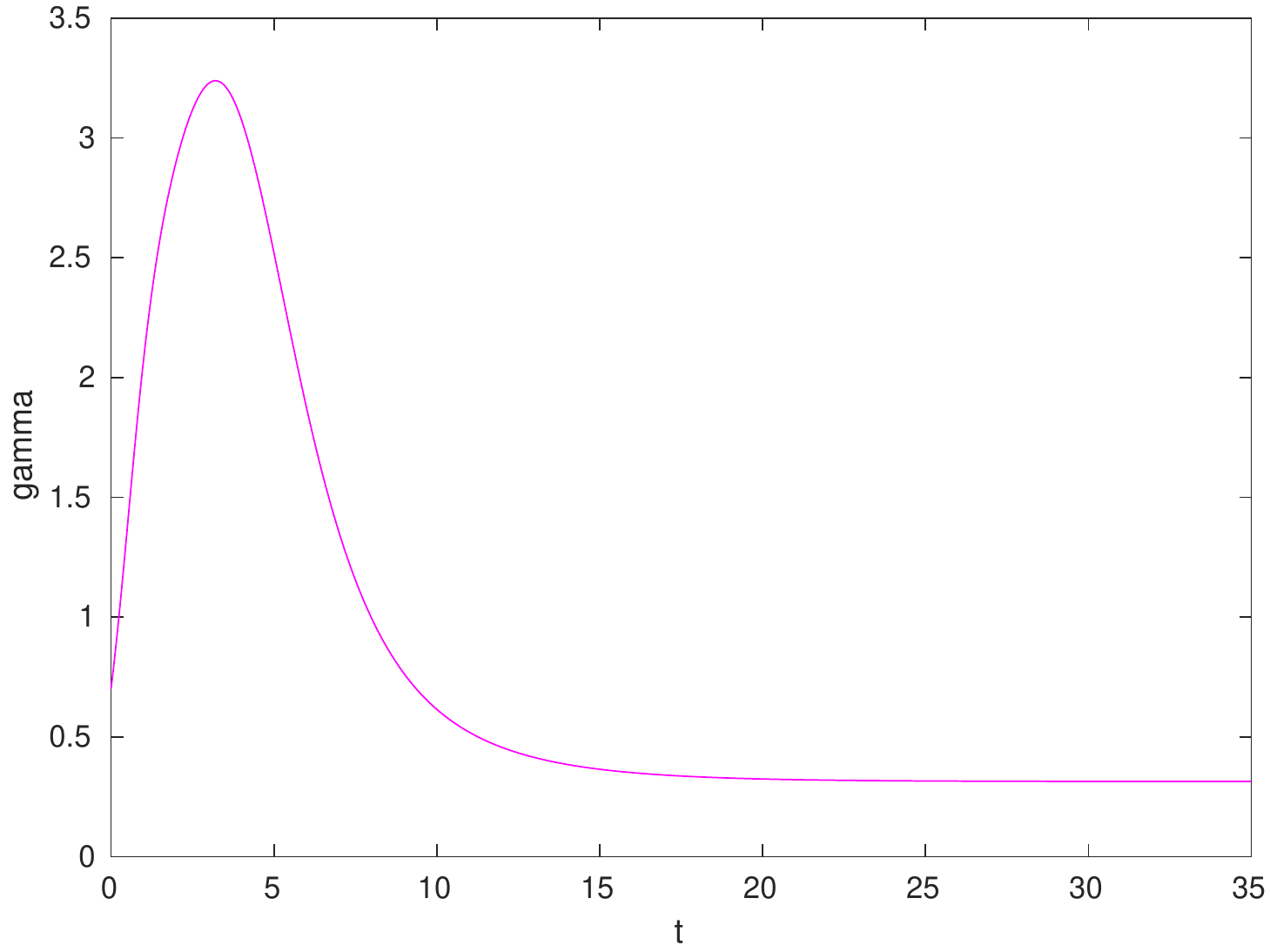}
    \caption{Typical form of time-dependent $\gamma$ matching the SIR system to the eRG one.}
    \label{fig:gammat}
    \end{center}
    \end{figure}

In the original eRG framework the number of recovered individuals were not explicitly taken into account. This is, however, straightforward to implement by  introducing an equation for $dR/dt$ and imposing a conservation law
equivalent to the one for the SIR model. A minimal choice compatible with the conservation law is  
 \begin{eqnarray} 
 \label{modifiedSIR}
  \frac{dR}{dt} &=& \varepsilon I \;, \nonumber \\
  \frac{dI}{dt} &=& \tilde{\gamma}(I+R) \ln(I+R) \left( 1- \frac{\ln(I+R)}{\ln P}\right) - \varepsilon I \;, \\
    \frac{dS}{dt} &=& - \tilde{\gamma}(I+R) \ln(I+R) \left( 1- \frac{\ln(I+R)}{\ln P}\right) \nonumber \;,
\end{eqnarray}
where the parameters are $\tilde{\gamma}$, $P$ and $\varepsilon$.
  At fixed $N$, $\tilde{\gamma}$ , $P$ and for any value of $\varepsilon$, the SIR model in \eqref{SESE} and the eRG systems of equations match  if  we allow $\gamma$ to be the following time-dependent function \be
  \gamma(t) = \tilde{\gamma}(I(t)+R(t)) \ln (I(t) + R(t)) \left( 1- \frac{\ln(I(t)+R(t))}{\ln P}\right) \frac{N}{I(t) S(t)}\;.
  \ee
  As we shall see this is a welcome feature. To better appreciate the mapping we show in figure~\ref{fig:gammat}  the time-dependent $\gamma$ parameter for a hypothetical case
  with $N= 7$ millions, $P=300$ thousands, $\varepsilon=1$ and $\tilde{\gamma}=0.7$ with initial conditions $I(0)=3$, $R(0)=0$ and $S(0)=N-I(0)$.
The result is a smooth function that peaks at short times and then plateaus to a fraction of $\tilde{\gamma}$. In other words the eRG naturally encodes a rapid diffusion of the disease in the initial states of the epidemic and the slow down at later times. 

\subsection{Reproduction number}
An important quantity for pandemics is the reproduction number related  to the expected average number of infected cases due to one  case. In the time-dependent generalised SIR model it is identified as: 
\begin{eqnarray}
R_0(t)= \frac{\gamma(t)} {\varepsilon(t)} \ .
\end{eqnarray}
  To extract $R_0$ from data it is useful to recast it as: 
 \begin{equation}
 \label{videomare}
    R_0=\frac{\gamma}{\varepsilon}=  \frac{\frac{dI}{dt}+ \frac{dR}{dt}}{\frac{dR}{dt}} \frac{N}{S} =
    \frac{\frac{d\tilde{I}}{dt}}{\frac{dR}{dt}}  \frac{N}{S} \;.
  \end{equation}
  The result holds for $\gamma$ and $\varepsilon$ generic functions of time. Here we also generalise the time-dependence of $\varepsilon$ to be:  
 \be
 \label{eps}
\varepsilon(t) = A\left[1- c\cdot e^{-\frac{1}{2}\left( \frac{t-t_0}{W}\right)^2} \right]\;.
\ee
As it is clear from its form this function has a dip at $t_0$ (possibly correlated with the peak of the newly infected cases) of width $W$  and depth $c \cdot A$ with $A$ the asymptotic value for  $t\to \pm \infty$. 
  
    As we shall see the shape allows for a substantial increase of $R_0$ near the peak of the newly infected cases that could be due to a number of factors including possible health-system stress around this period. 
 
\section{Testing the framework}
As a timely application we consider the COVID-19 pandemic. Here the factor $N/S$ in \eqref{videomare} can be neglected  as the number of total infected is at most of O$(1\%)$ of the total susceptible population and the ratio is therefore very close to unity.
  The reproduction number can hence be estimated as:
  \be
  R_0(t) =\frac{\rm newly \; infected(t)}{\rm newly \; recovered(t)} \;.
  \ee
The values for the numerator and denominator for different regions of the world can be obtained from several sources such as the  \href{https://www.who.int/csr/sars/country/en/}{World Health Organization (WHO)} and   \href{https://www.worldometers.info/coronavirus/}{Worldometers}.
   
   As testbed scenarios we consider four benchmark cases,  namely  Denmark, Germany, France and Italy. These countries adopted different degrees of containment measures. 
  We find convenient to  bin the data in weeks  to smooth out daily fluctuations. We associate an error
  to both newly infected and newly recovered given by the square root of their values. 
  
  Procedurally we first fit the function $\alpha(t)$ to determine $a$, $b$ and $\gamma$ following \cite{DellaMorte:2020wlc,Cacciapaglia:2020mjf}. We then solve with these, as input, the system of equations \eqref{modifiedSIR} and $\varepsilon(t)$ given in \eqref{eps}. The parameters entering in the function $\varepsilon(t)$ are obtained by performing a $\chi^2$ minimization to the data related to the recovered cases. Combining the results with $\gamma(t)$ 
we compare  $R_0(t)$ with the actual data. 
 
 For each country we show the data and the model results by grouping together five graphs in a single figure. The different panels represent $\tilde{I}(t)$, $R(t)$, $R_0 (t)$, $\gamma(t)$ and $\varepsilon (t)$, all as function of the week number.  Additionally the data will be reported starting some time after the outbreak. The reason being that the values of the number of recovered cases at early times is too small to be reliable and begins to be sizeable only few weeks after the outbreak. 
 For our predictions $\varepsilon(t)$, $\gamma(t)$ and $R_0(t)$ we show bands limiting the 90\% confidence level. Those are obtained shifting the data for the number of recovered cases by 1.65 standard deviations. The fitting errors for $\tilde{\gamma}$, $a$ and $b$ from the method in \cite{DellaMorte:2020wlc} can be neglected given that these parameters are highly constrained by the data.

 In general we find good agreement between the data
 and the model with the exception of the increase in the number of infected cases  occurring in the last weeks for some of the countries.
 Those are non-smooth events resulting from an abrupt change in the social distancing measures or from new and previously un-accounted disease hotspots.
 As such those events cannot be predicted by smooth models.

  \subsection{Denmark}
\begin{figure}
    \centering
    \begin{subfigure}[t]{0.32\textwidth}
        \centering
        \includegraphics[width=\linewidth]{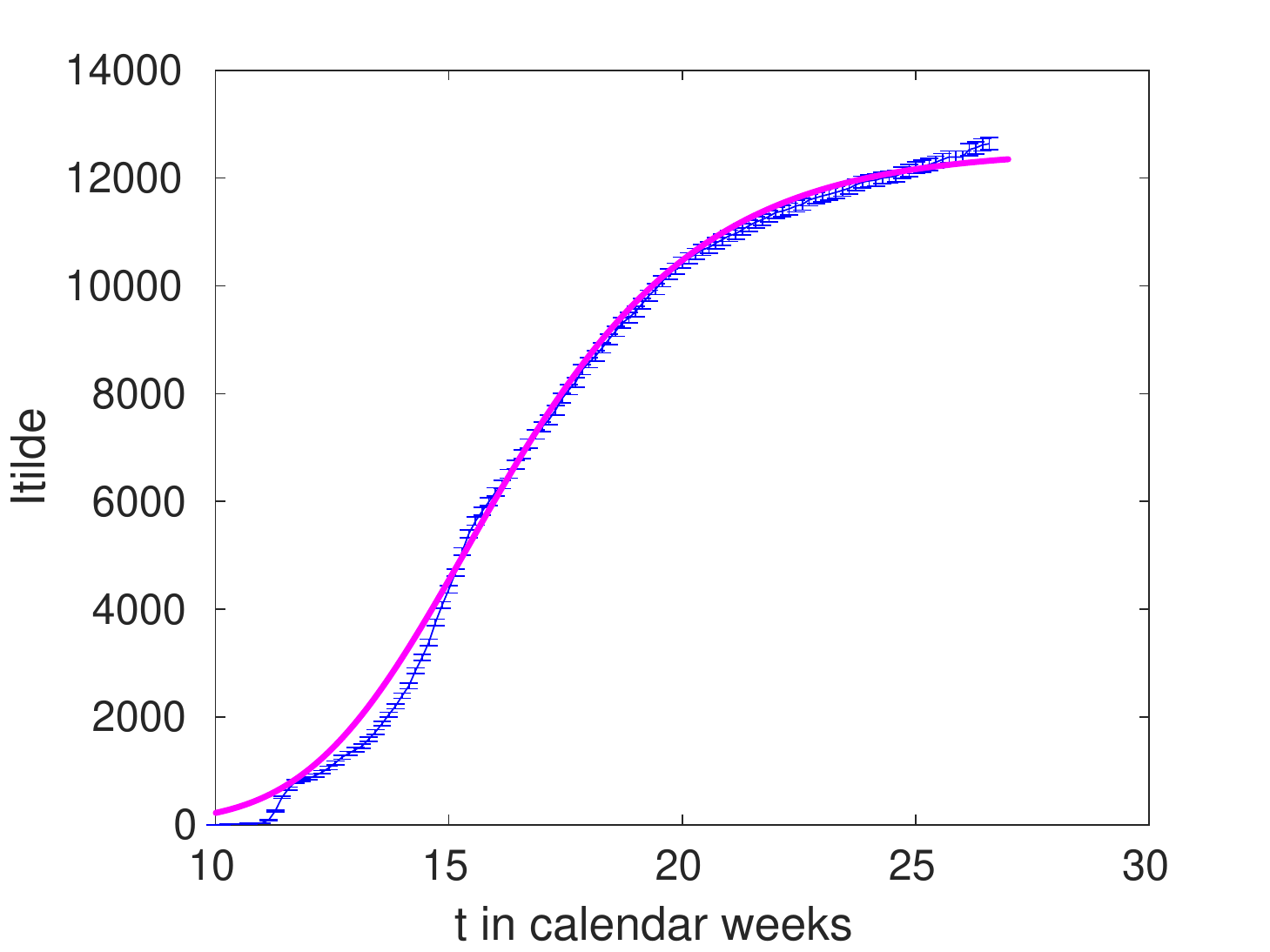} 
        \caption{Cumulative number of infected cases $\tilde{I}$.} \label{Itilde-DK}
    \end{subfigure}
     \begin{subfigure}[t]{0.32\textwidth}
        \centering
        \includegraphics[width=\linewidth]{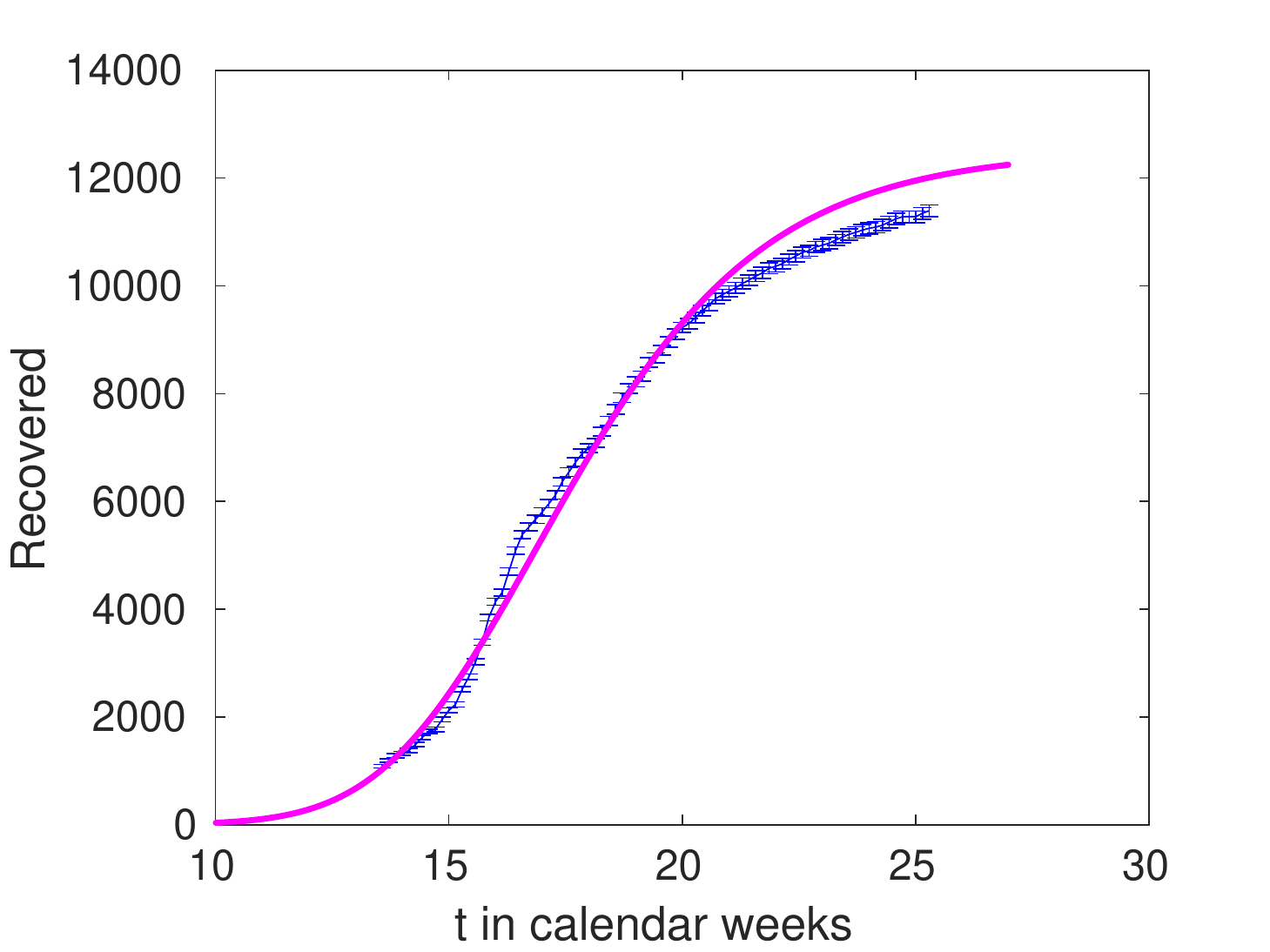} 
        \caption{Number of recovered cases.} \label{R-DK}
    \end{subfigure}
\begin{subfigure}[t]{0.32\textwidth}
        \centering
        \includegraphics[width=\linewidth]{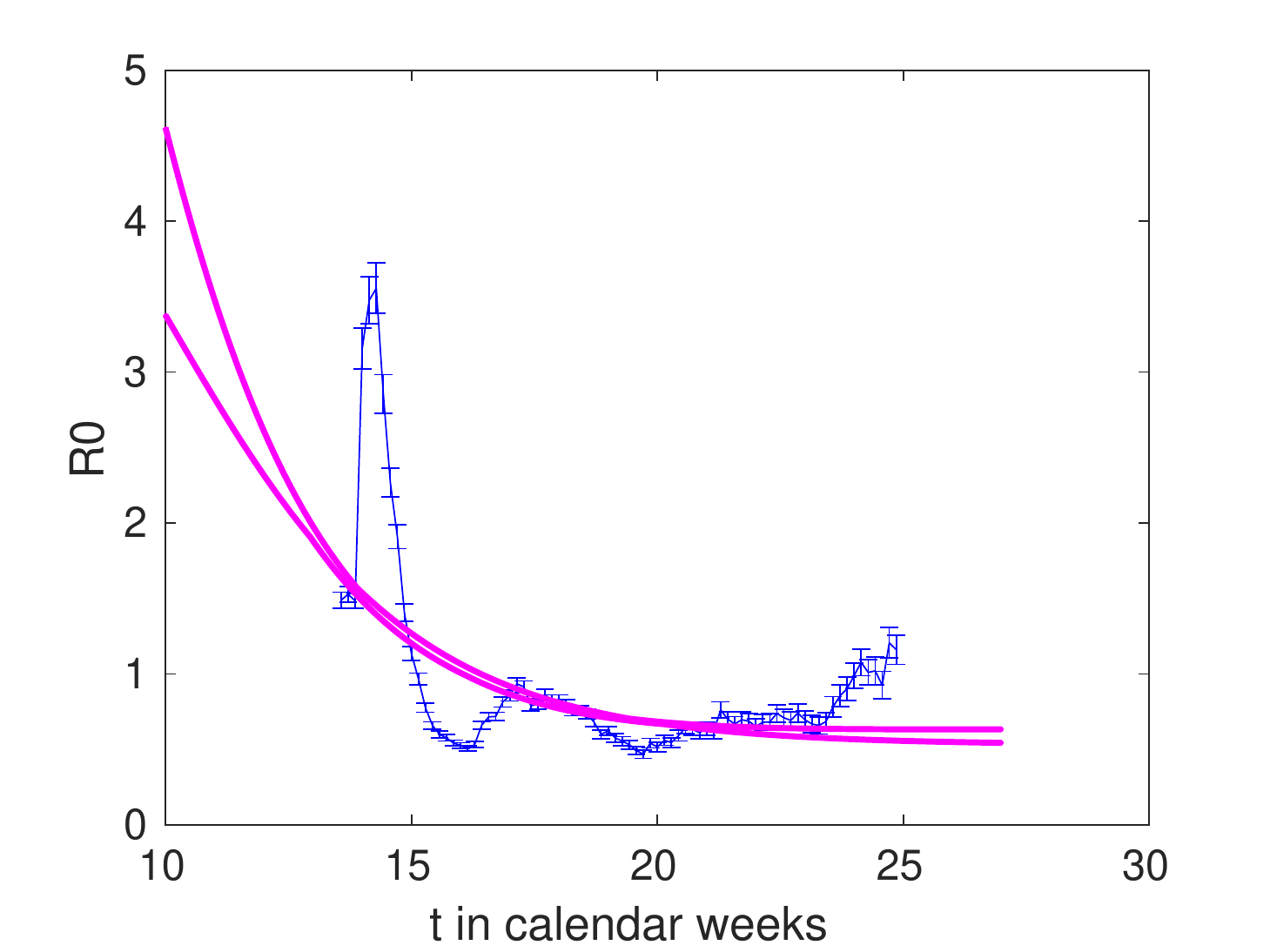} 
        \caption{Reproduction number.} \label{R0-DK}
    \end{subfigure}

    \vspace{1cm}
    \begin{subfigure}[t]{0.35\textwidth }
    \centering
        \includegraphics[width=\linewidth]{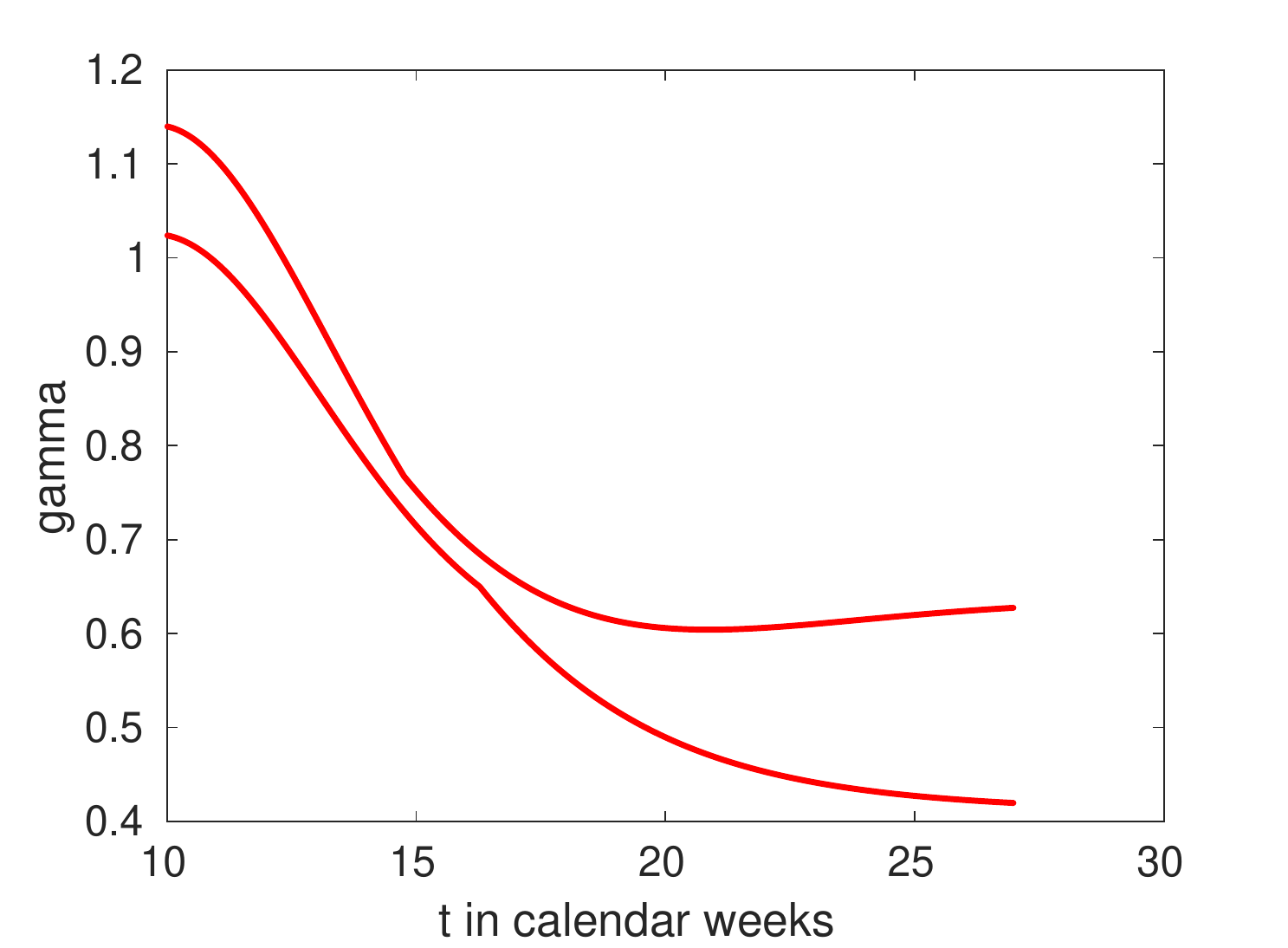} 
        \caption{ $\gamma(t)$} \label{gamma-DK}
    \end{subfigure}
     \hskip 1cm
     \begin{subfigure}[t]{0.35\textwidth }
    \centering
        \includegraphics[width=\linewidth]{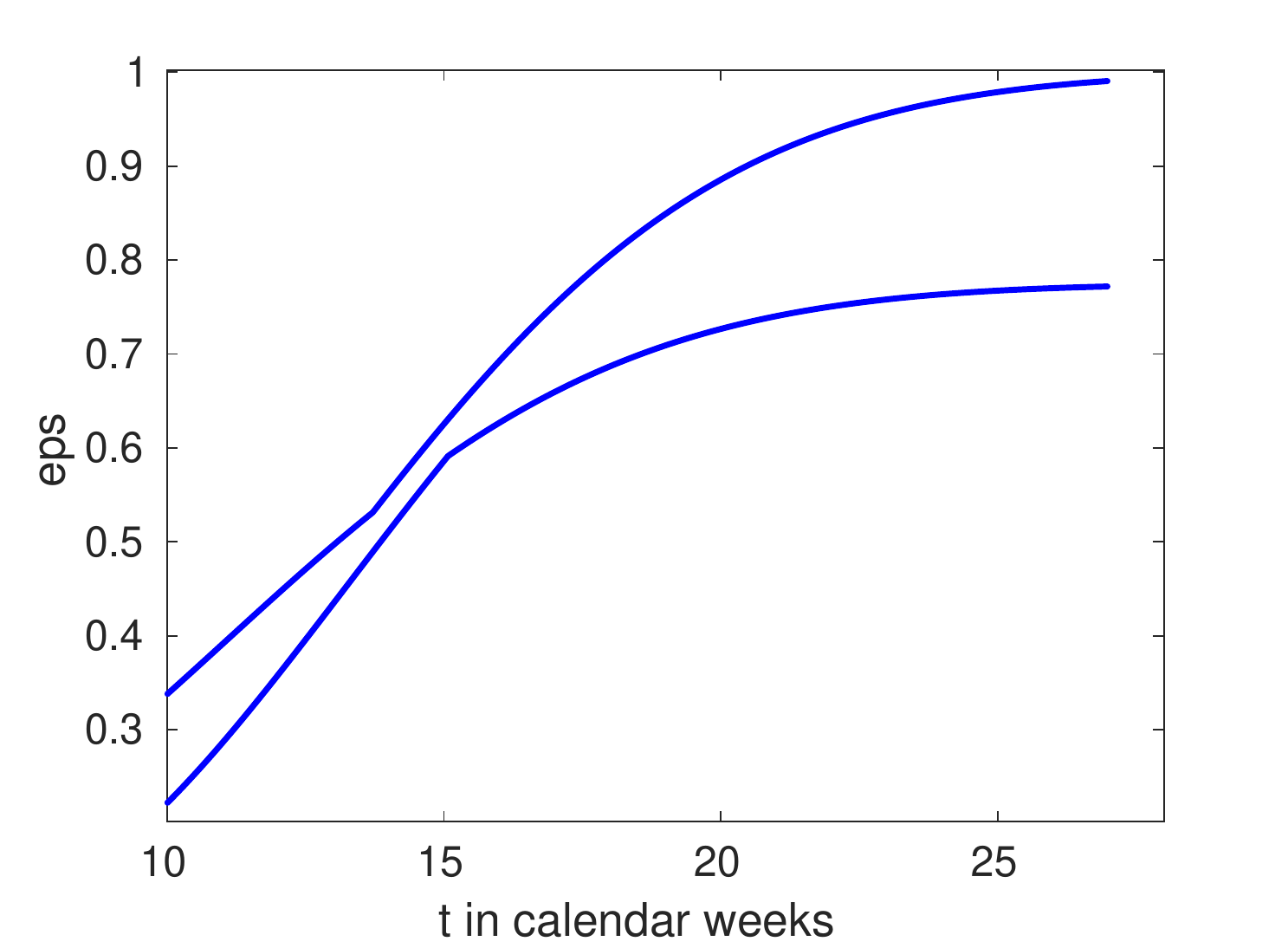} 
        \caption{Recovery rate $\varepsilon(t)$} \label{eps-DK}
    \end{subfigure}
    \caption{Time dependence of $\tilde{I}(t)$, $R(t)$, $R_0(t)$, $\gamma(t)$ and $\varepsilon(t)$ within the eRG SIR model optimised to describe the Danish data for COVID-19. Solid lines are the model (shown as a 95\% confidence level band for the predictions) and the dots with error-bars are the data.  For the chosen confidence intervals the errors for the fit functions $\tilde{I}$ and ${R}$ are not visible in this and the following figures.}
    \label{Danimarca}
\end{figure}

The data and the model results for Denmark are shown in Fig.~\ref{Danimarca}.  For the time-dependence of $R_0$, given  in panel~\ref{R0-DK}, we observe that the model captures the variation of the data for over 10 weeks. Overall we find that the eRG model provides a reasonable description of the time dependence of the reproduction number. We observe that the recovery rate $\varepsilon(t)$ grows with time  by a factor of five. There could be several factors contributing to this growth, one being a better trained health system. 

\subsection{Germany}
\begin{figure}
    \centering
    \begin{subfigure}[t]{0.32\textwidth}
        \centering
        \includegraphics[width=\linewidth]{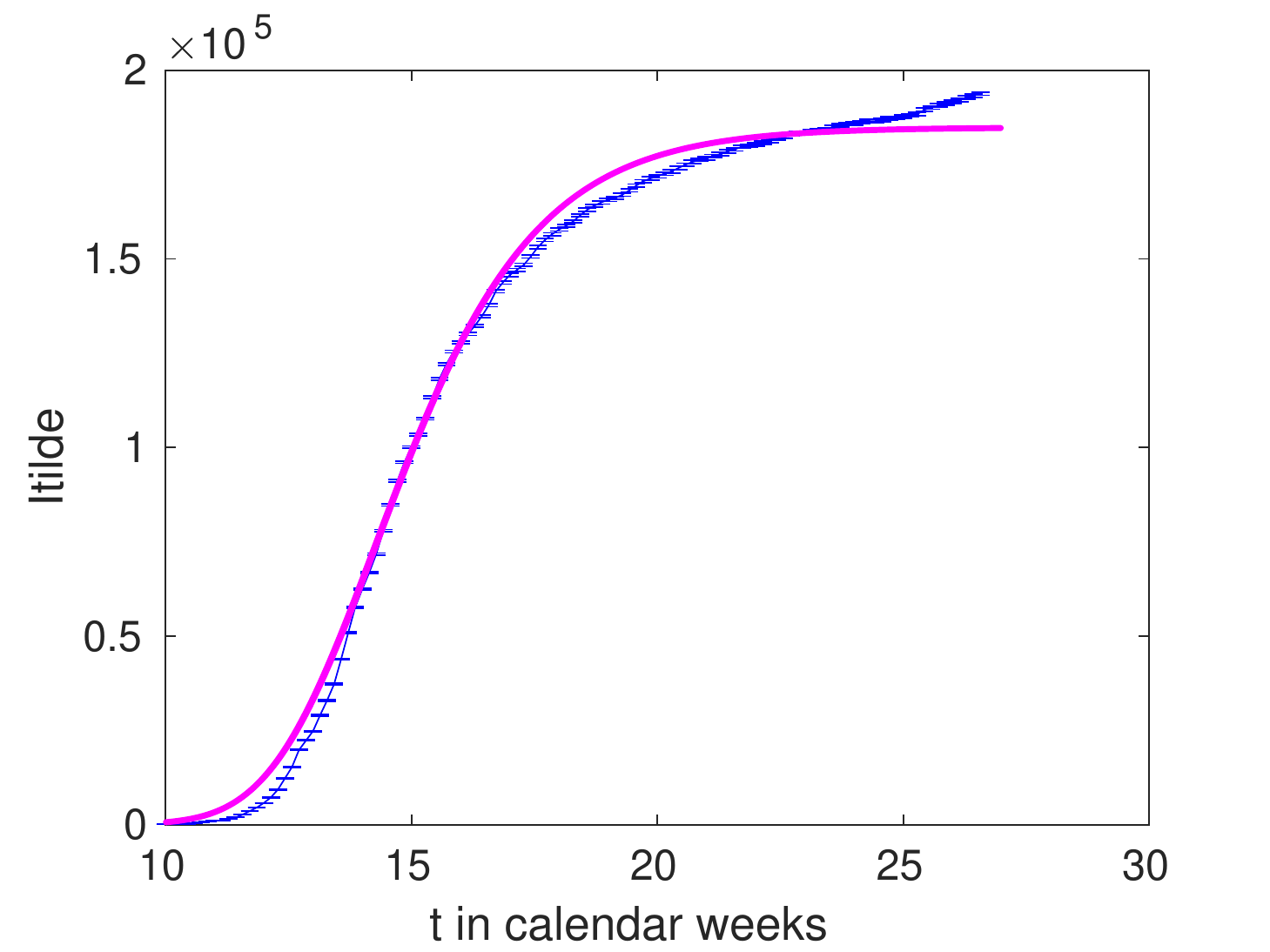} 
        \caption{Cumulative number of infected cases $\tilde{I}$.} \label{fig:timing1}
    \end{subfigure}
     \begin{subfigure}[t]{0.32\textwidth}
        \centering
        \includegraphics[width=\linewidth]{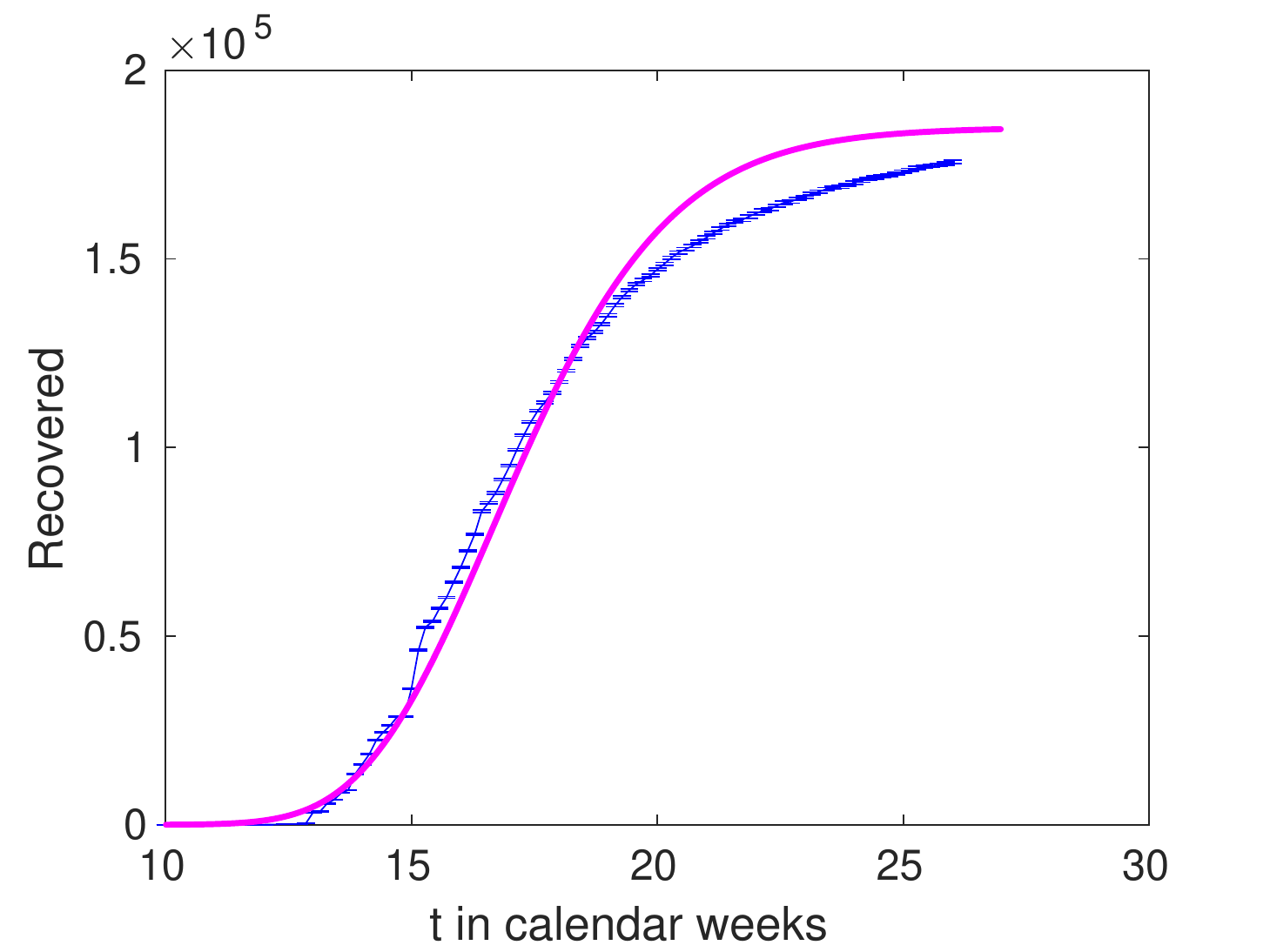} 
        \caption{Number of recovered cases.} \label{fig:timing2}
    \end{subfigure}
\begin{subfigure}[t]{0.32\textwidth}
        \centering
        \includegraphics[width=\linewidth]{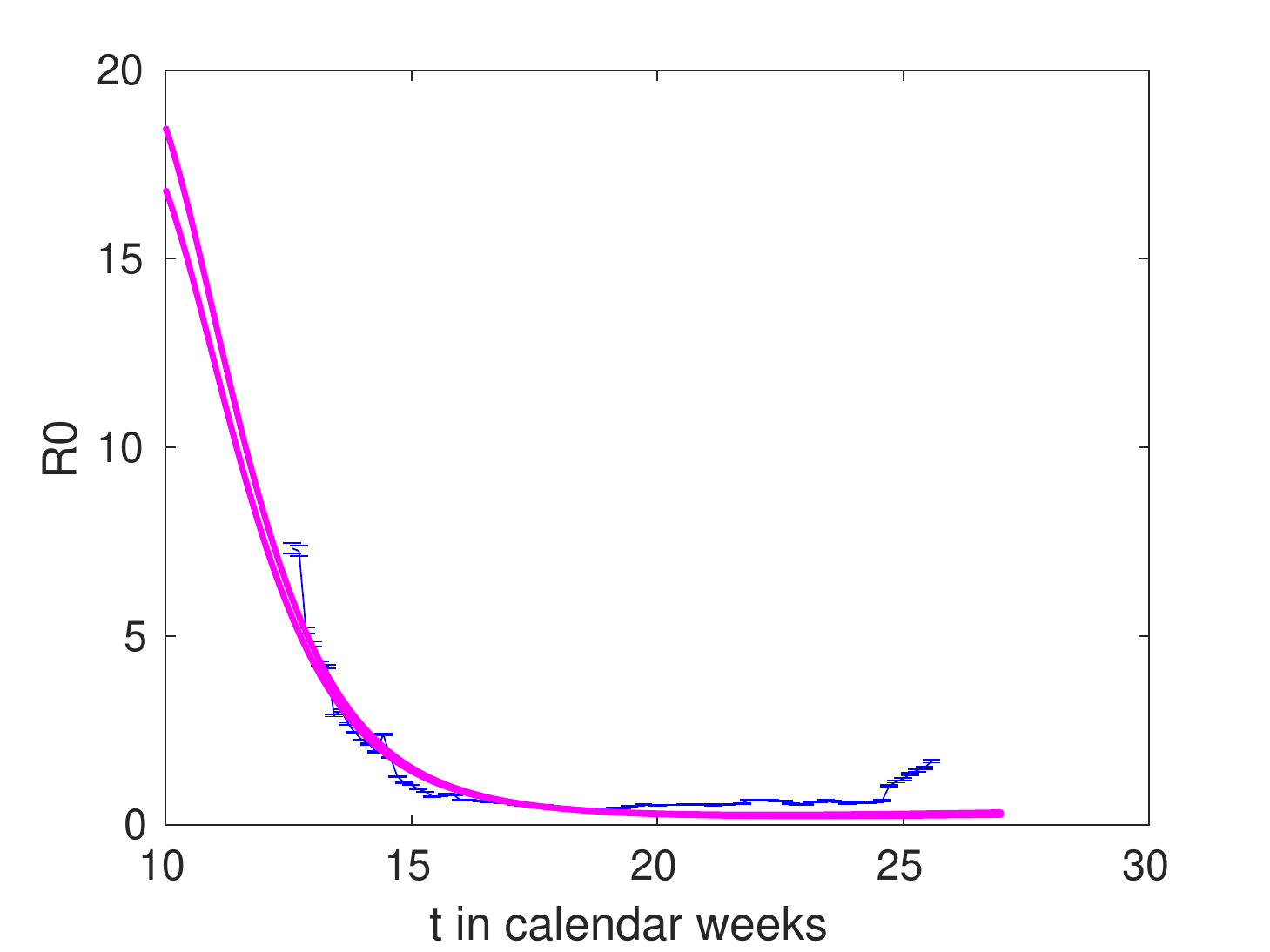} 
        \caption{Reproduction number.} \label{fig:timing2}
    \end{subfigure}

    \vspace{1cm}
    \begin{subfigure}[t]{0.35\textwidth }
    \centering
        \includegraphics[width=\linewidth]{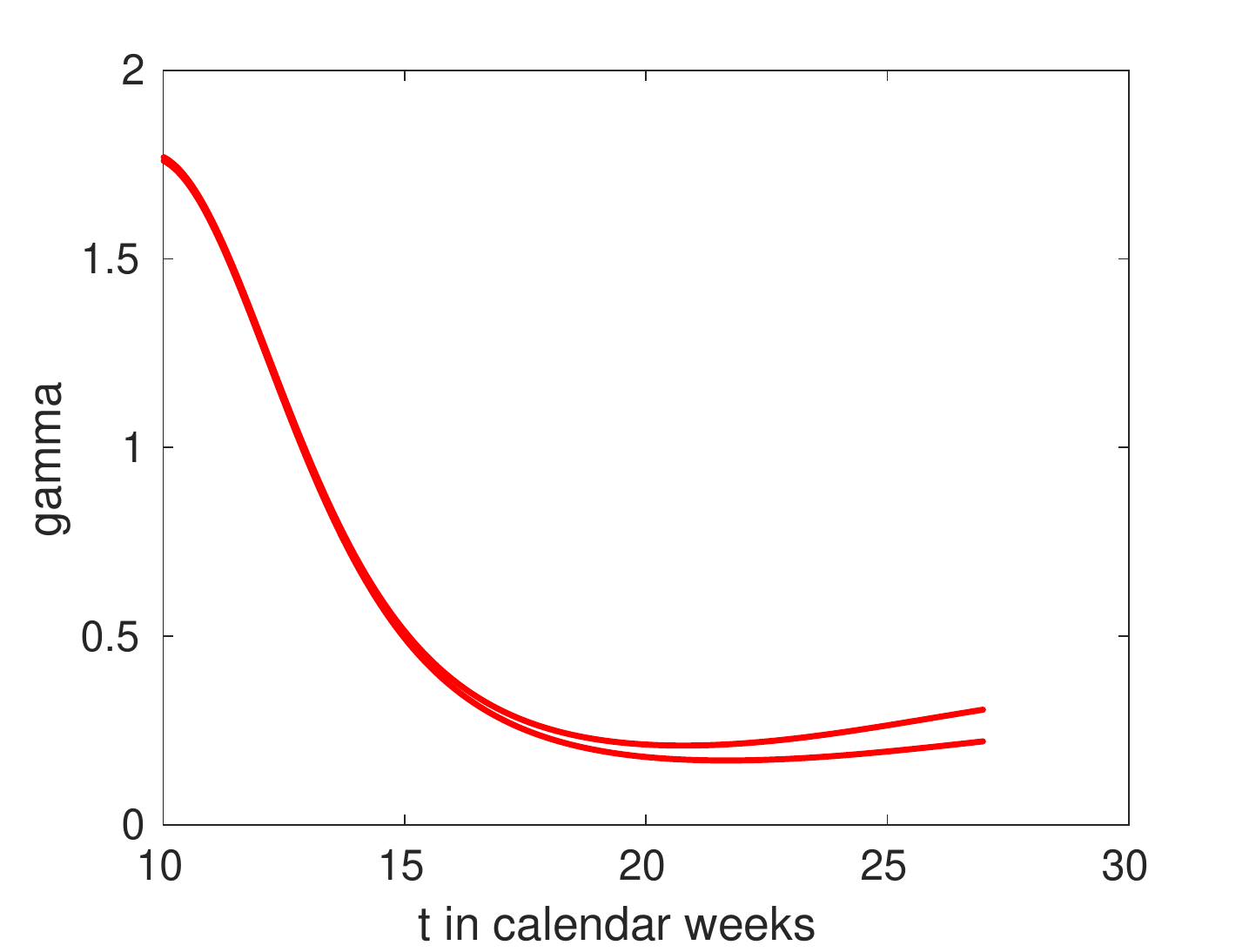} 
        \caption{ $\gamma(t)$} \label{fig:timing3}
    \end{subfigure}
     \hskip 1cm
     \begin{subfigure}[t]{0.35\textwidth }
    \centering
        \includegraphics[width=\linewidth]{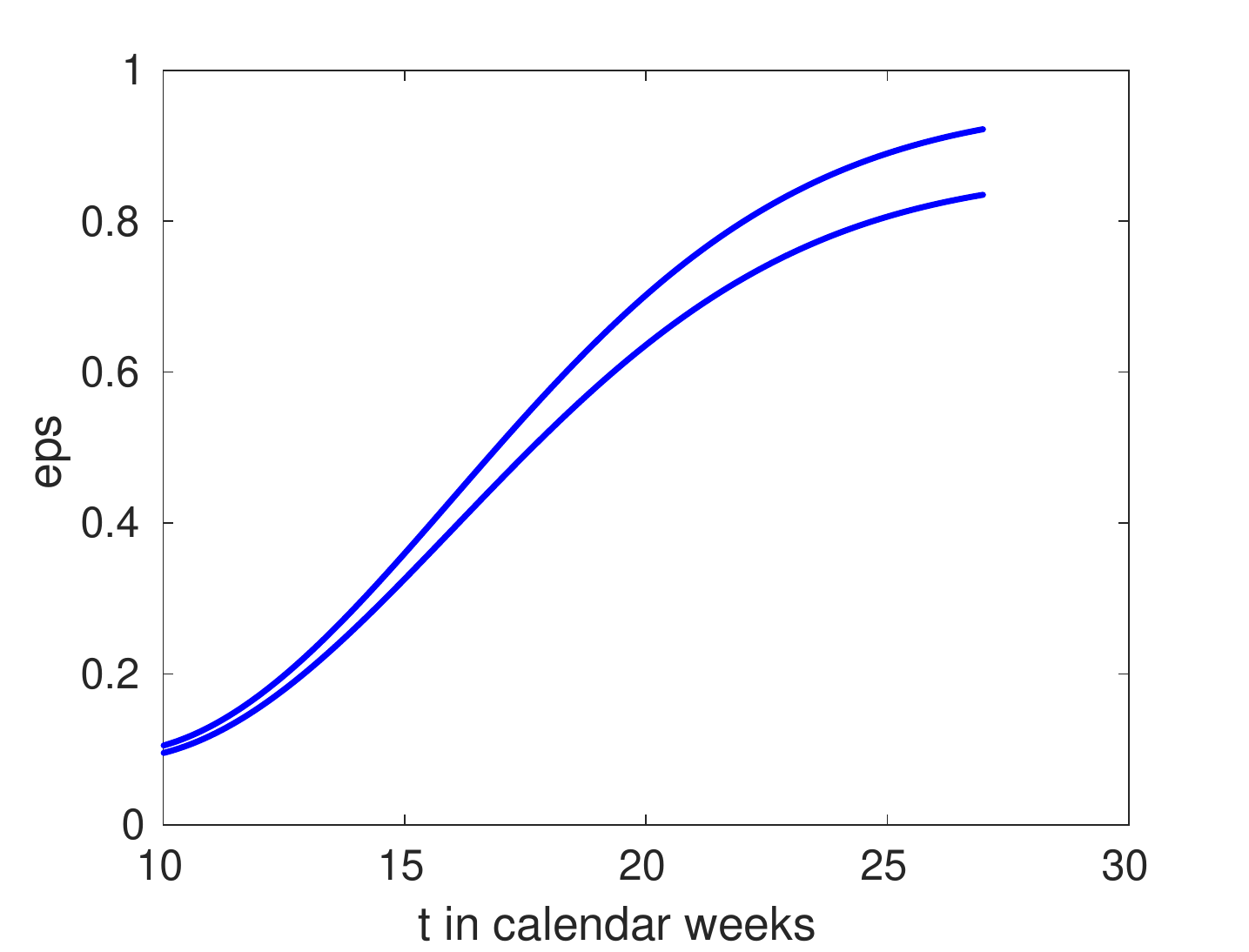} 
        \caption{Recovery rate $\varepsilon(t)$} \label{fig:timing3}
    \end{subfigure}
    \caption{Time dependence of $\tilde{I}(t)$, $R(t)$, $R_0(t)$, $\gamma(t)$ and $\varepsilon(t)$ within the eRG SIR model optimised to describe the German data for COVID-19. Solid lines are the model (shown as a 95\% confidence level band for the predictions) and the dots with error-bars are the data. }
    \label{Germania}
\end{figure}

For Germany the analysis is summarised in Fig.~\ref{Germania}.  The overall trends are similar to the Danish case including the temporal trend of the recovery rate $\varepsilon(t)$.

 \subsection{Italy}
\begin{figure}
    \centering
    \begin{subfigure}[t]{0.32\textwidth}
        \centering
        \includegraphics[width=\linewidth]{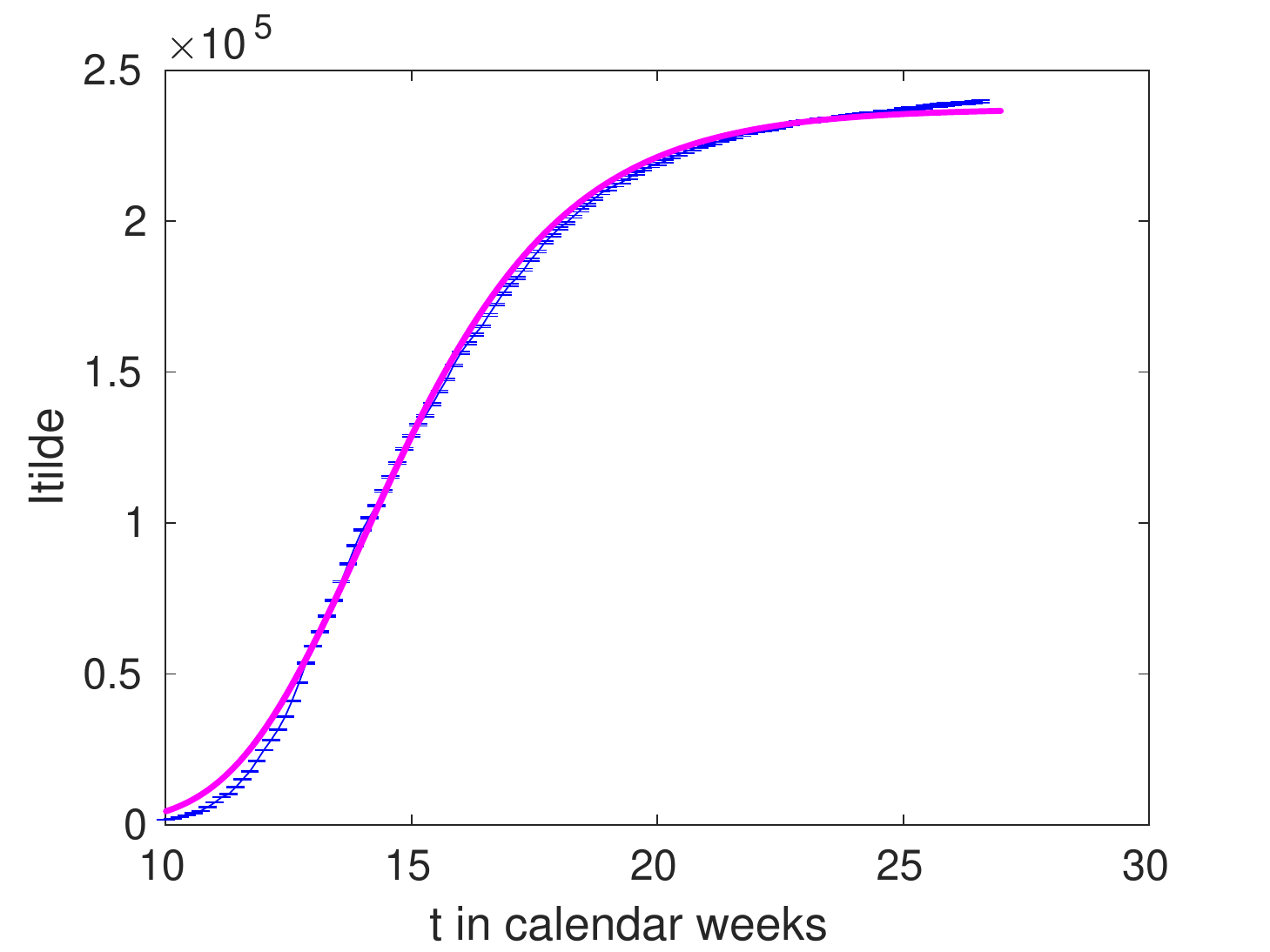} 
        \caption{Cumulative number of infected cases $\tilde{I}$.} \label{fig:timing1}
    \end{subfigure}
     \begin{subfigure}[t]{0.32\textwidth}
        \centering
        \includegraphics[width=\linewidth]{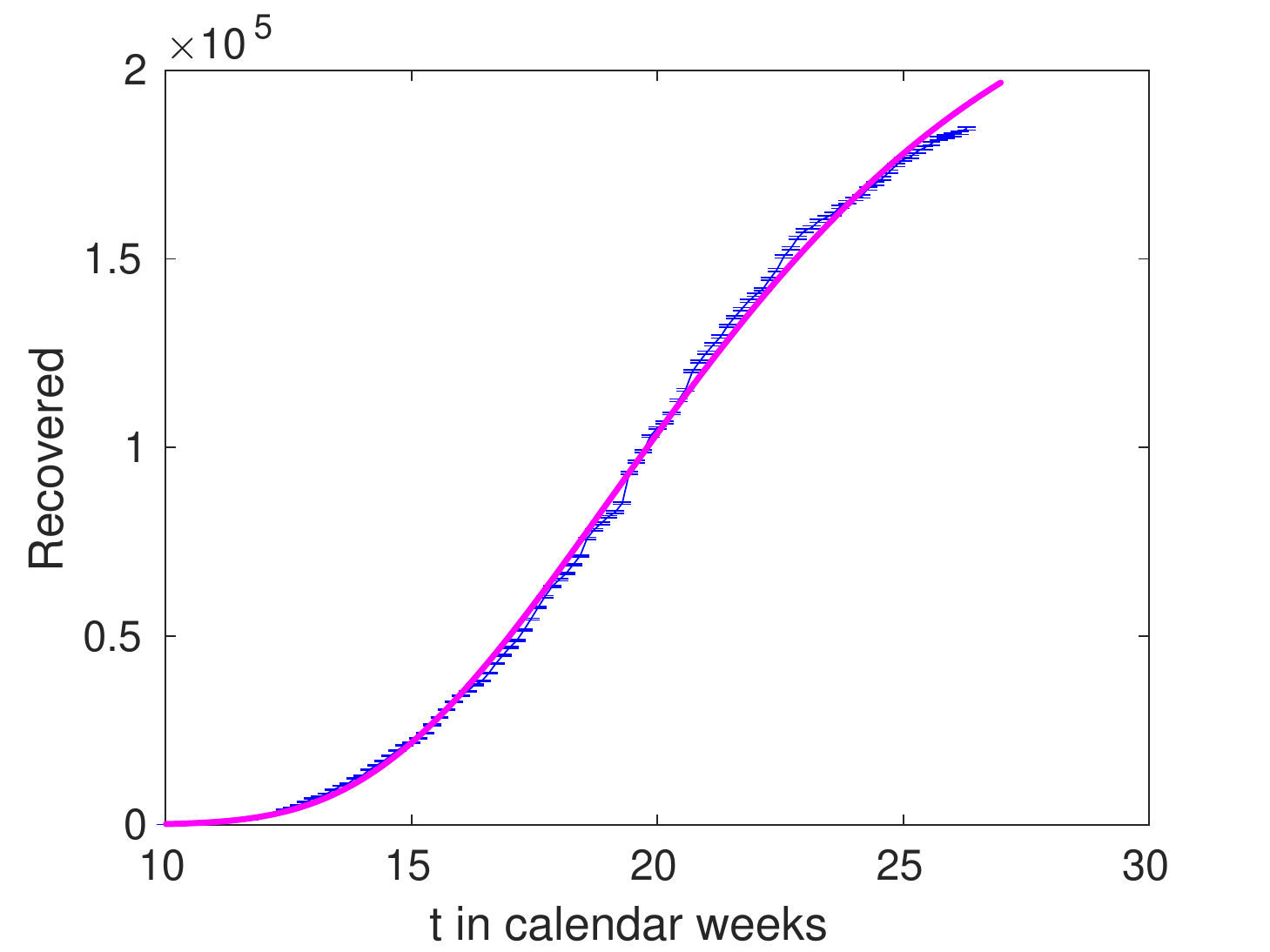} 
        \caption{Number of recovered cases.} \label{fig:timing2}
    \end{subfigure}
\begin{subfigure}[t]{0.32\textwidth}
        \centering
        \includegraphics[width=\linewidth]{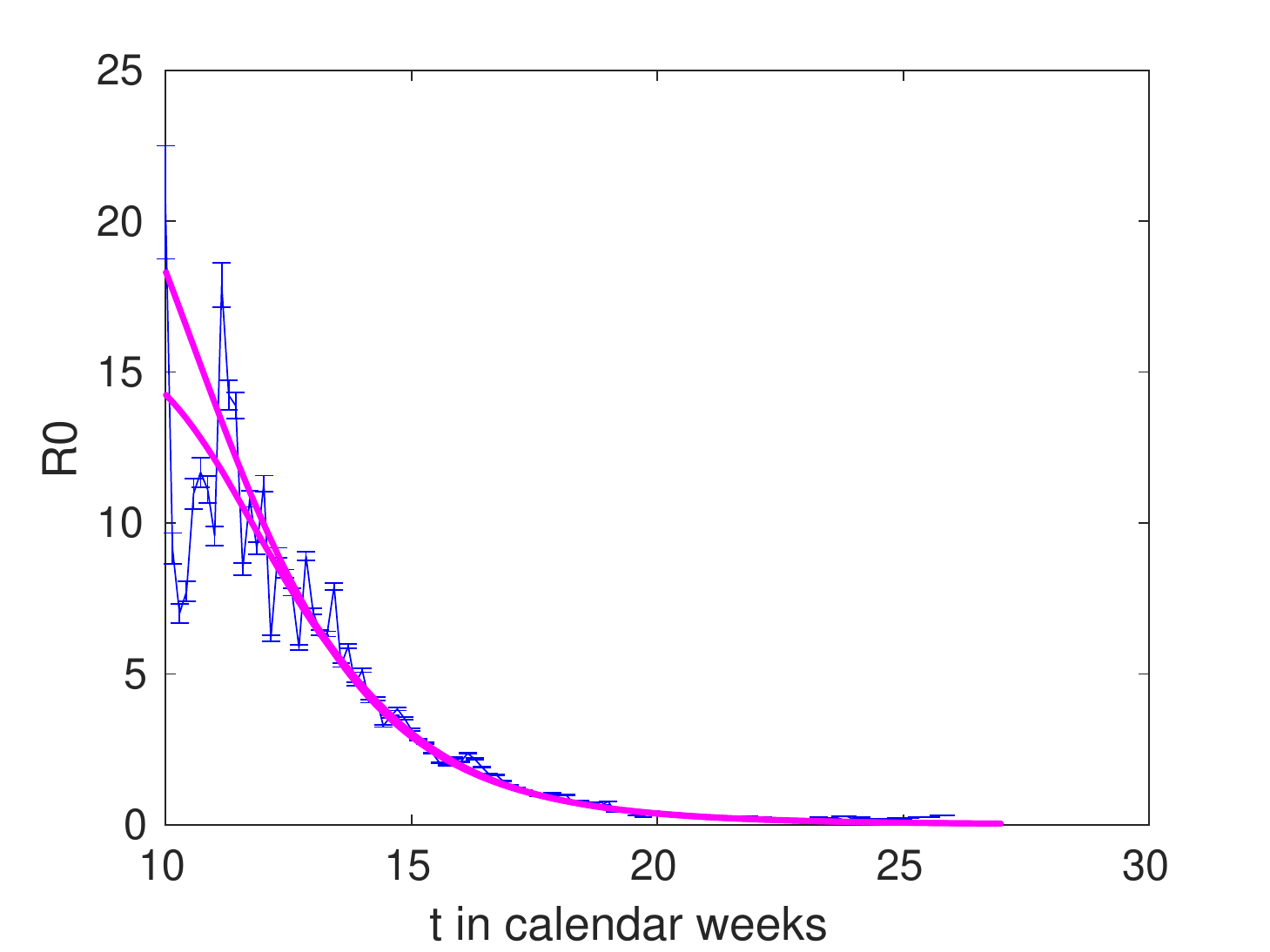} 
        \caption{Reproduction number.} \label{fig:timing2}
    \end{subfigure}

    \vspace{1cm}
    \begin{subfigure}[t]{0.35\textwidth }
    \centering
        \includegraphics[width=\linewidth]{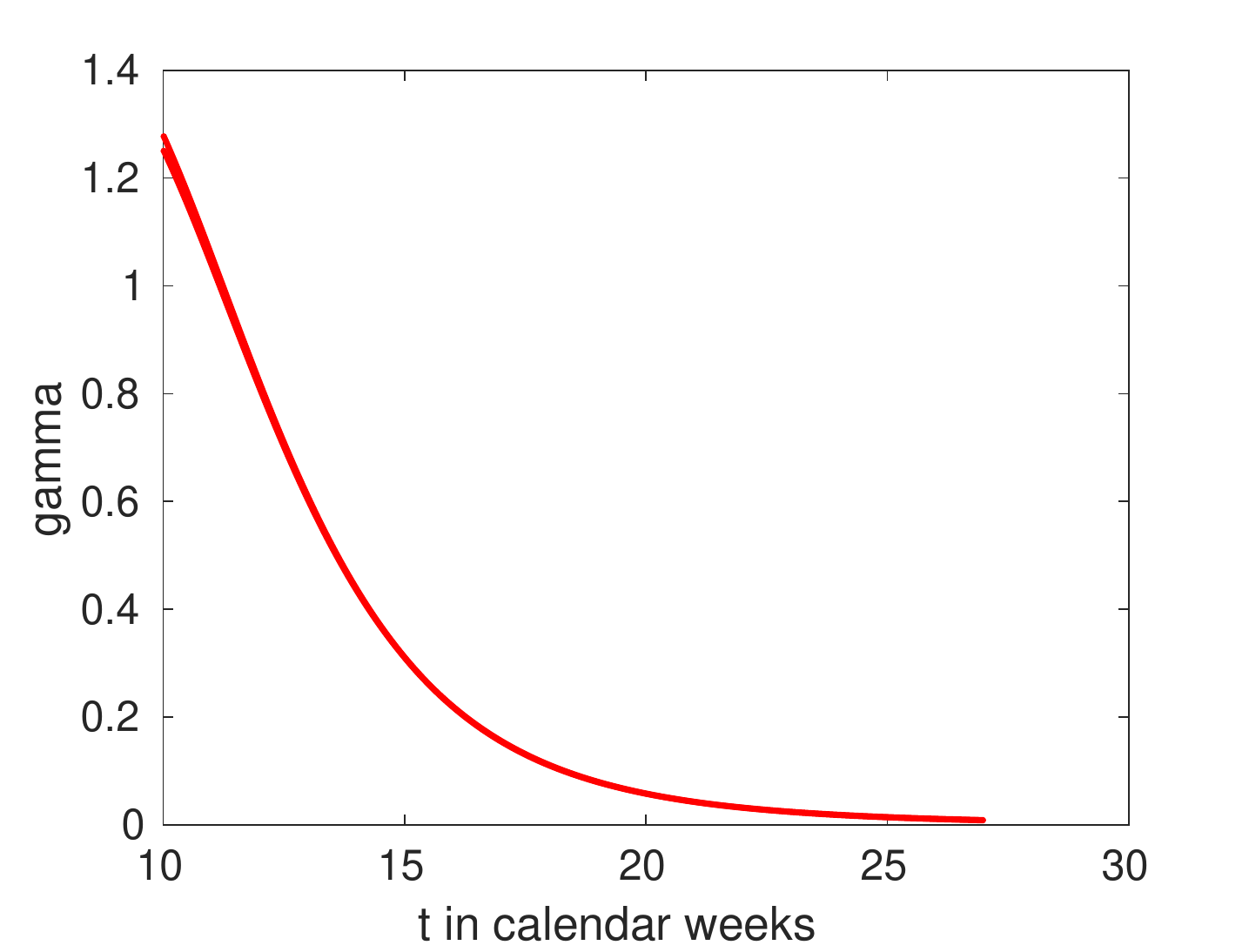} 
        \caption{ $\gamma(t)$} \label{fig:timing3}
    \end{subfigure}
     \hskip 1cm
     \begin{subfigure}[t]{0.35\textwidth }
    \centering
        \includegraphics[width=\linewidth]{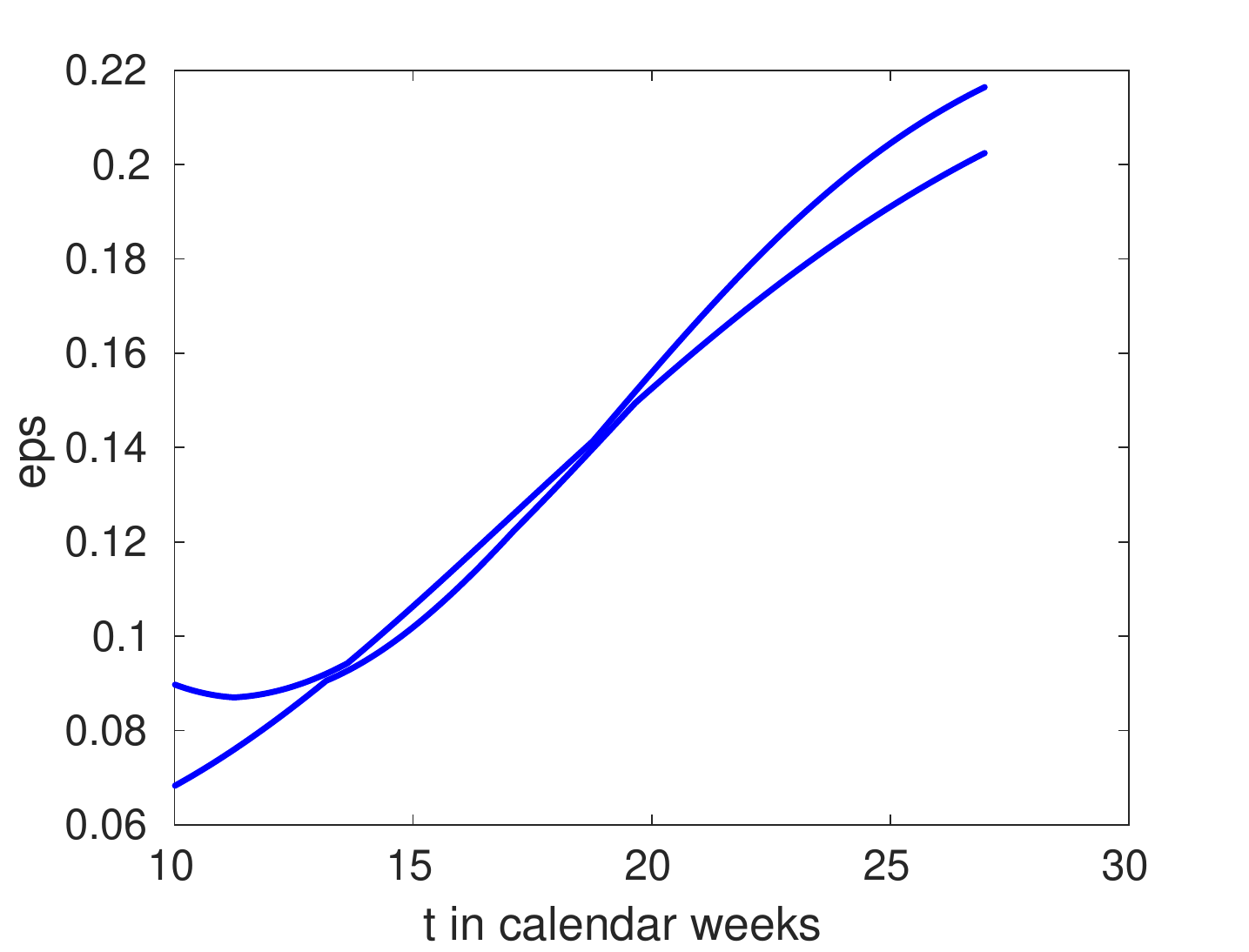} 
        \caption{Recovery rate $\varepsilon(t)$} \label{fig:timing3}
    \end{subfigure}
    \caption{Time dependence of $\tilde{I}(t)$, $R(t)$, $R_0(t)$, $\gamma(t)$ and $\varepsilon(t)$ within the eRG SIR model optimised to describe the Italian data for COVID-19. Solid lines are the model (shown as a 95\% confidence level band for the predictions)  and the dots with error-bars are the data.}
    \label{Italia}
\end{figure}

The analysis for Italy is shown in 
 figure~\ref{Italia}. We observe rather large values of $R_0$ compared to Denmark at early times and a factor of two with respect to Germany.  We also observe that a good fit is obtained for $\gamma(t)$ approaching very small values at large times. This is different from Germany and Denmark, suggesting strong distancing measures being adopted by the Italian government. This seems to be further followed by a smaller value of the recovery rate, roughly about a fourth.  However this last comparison is biased by the fact that the number of deaths in Italy is about 15\% of the number of infected cases while in Germany and Denmark it is below 5\% suggesting that a more accurate description at large times would require introducing also a compartment accounting for the deaths.

\subsection{France}
\begin{figure}
    \centering
    \begin{subfigure}[t]{0.32\textwidth}
        \centering
        \includegraphics[width=\linewidth]{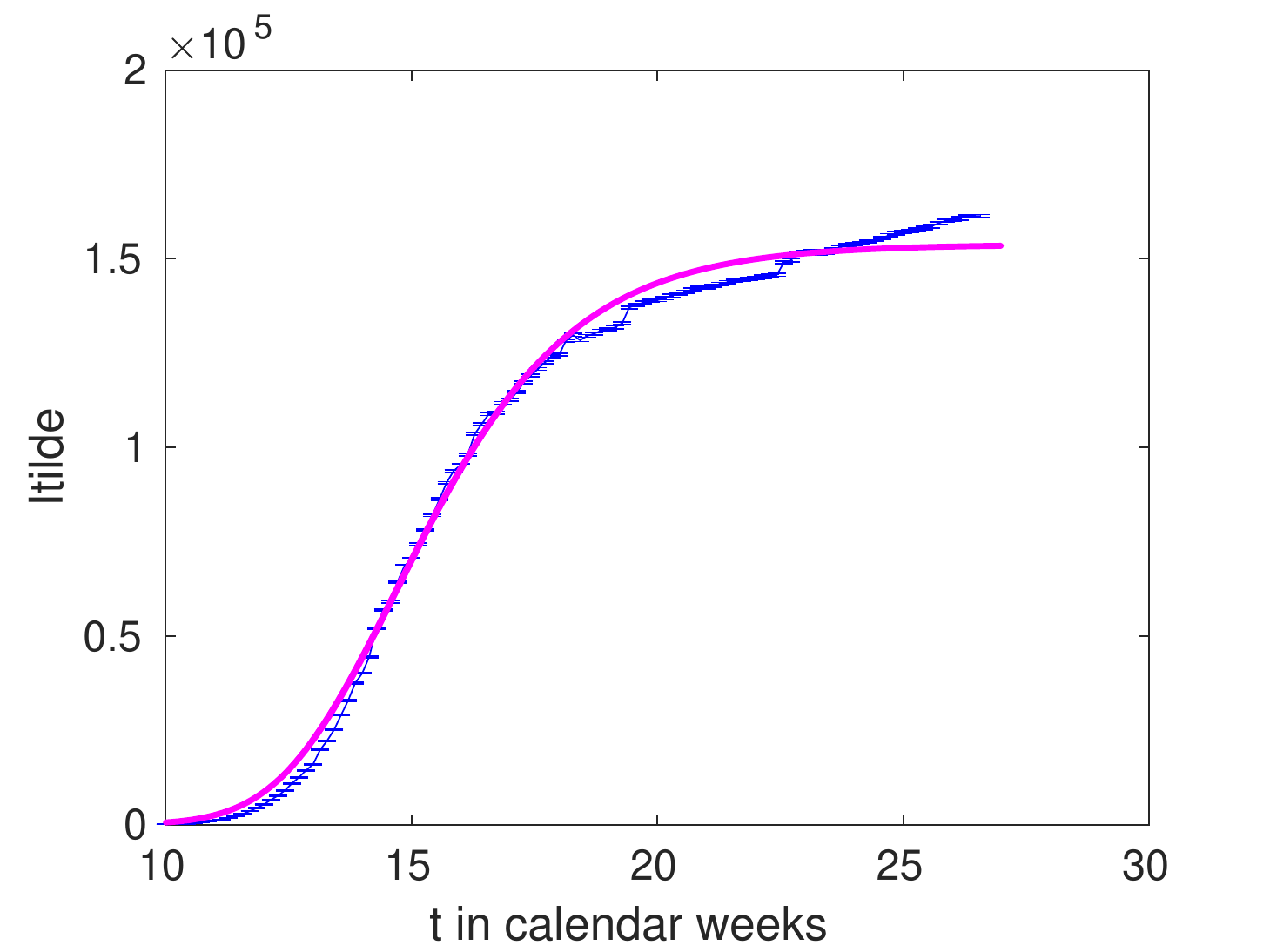} 
        \caption{Cumulative number of infected cases $\tilde{I}$.} \label{fig:timing1}
    \end{subfigure}
     \begin{subfigure}[t]{0.32\textwidth}
        \centering
        \includegraphics[width=\linewidth]{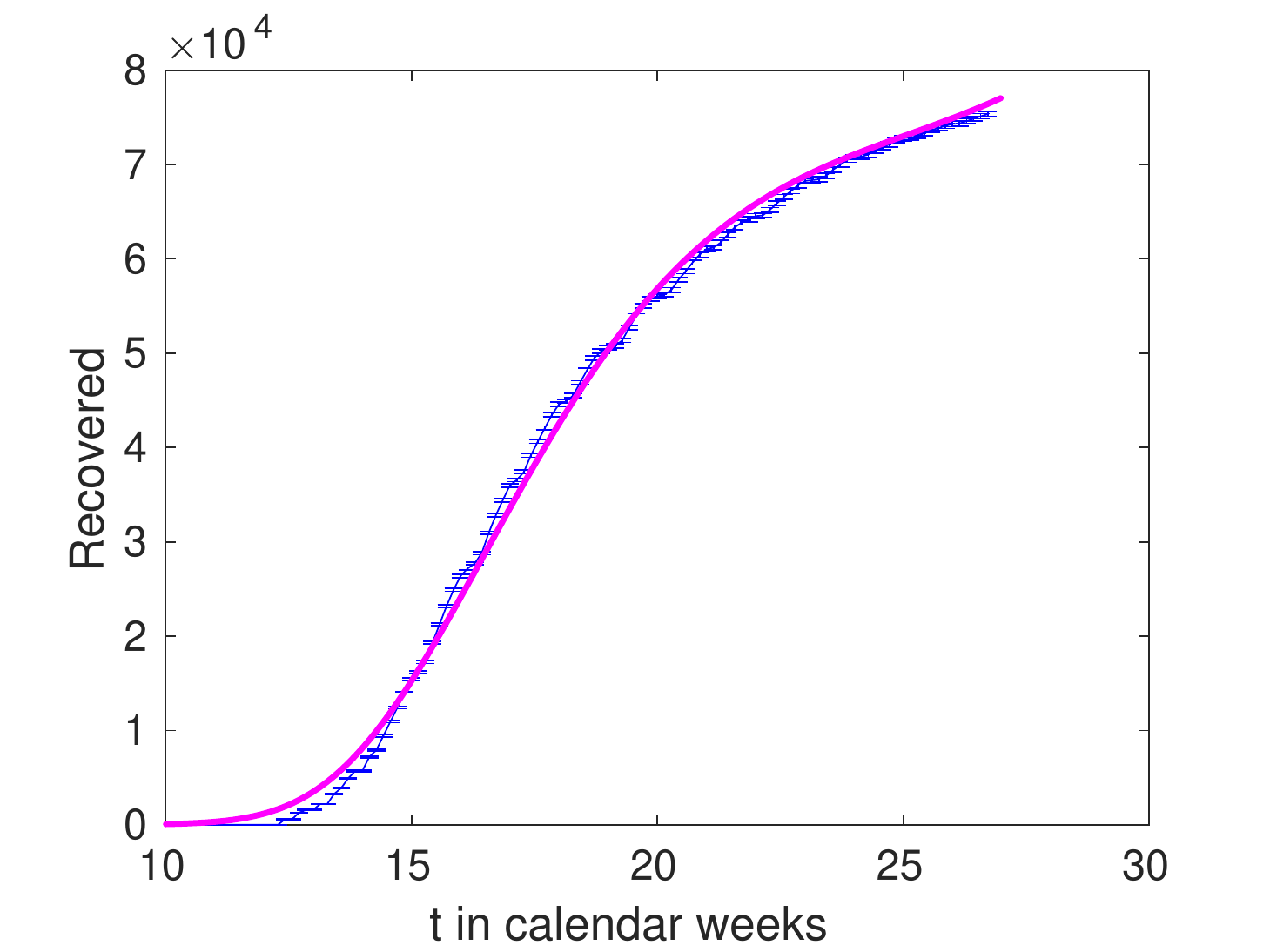} 
        \caption{Number of recovered cases.} \label{fig:timing2}
    \end{subfigure}
\begin{subfigure}[t]{0.32\textwidth}
        \centering
        \includegraphics[width=\linewidth]{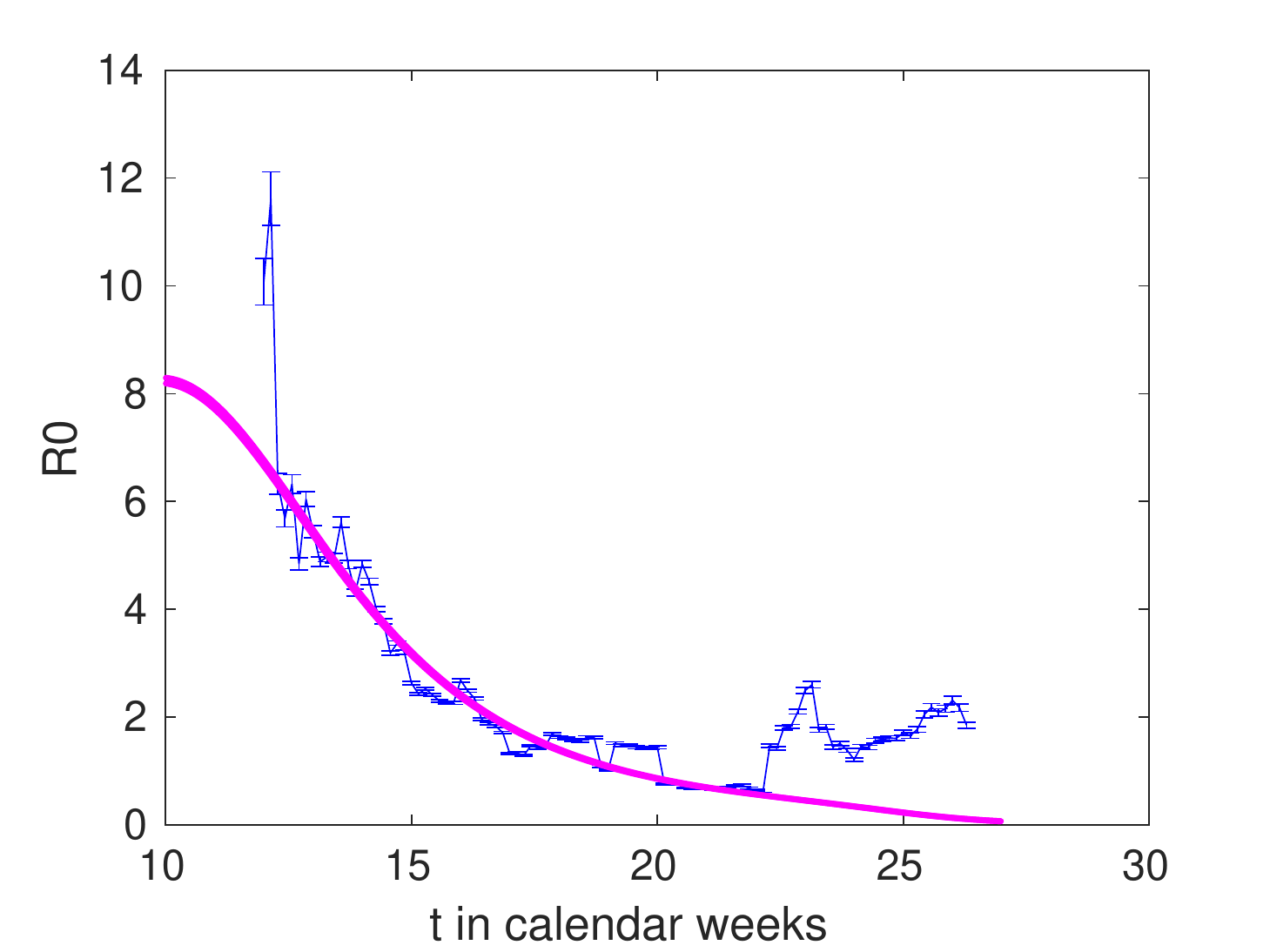} 
        \caption{Reproduction number.} \label{fig:timing2}
    \end{subfigure}

    \vspace{1cm}
    \begin{subfigure}[t]{0.35\textwidth }
    \centering
        \includegraphics[width=\linewidth]{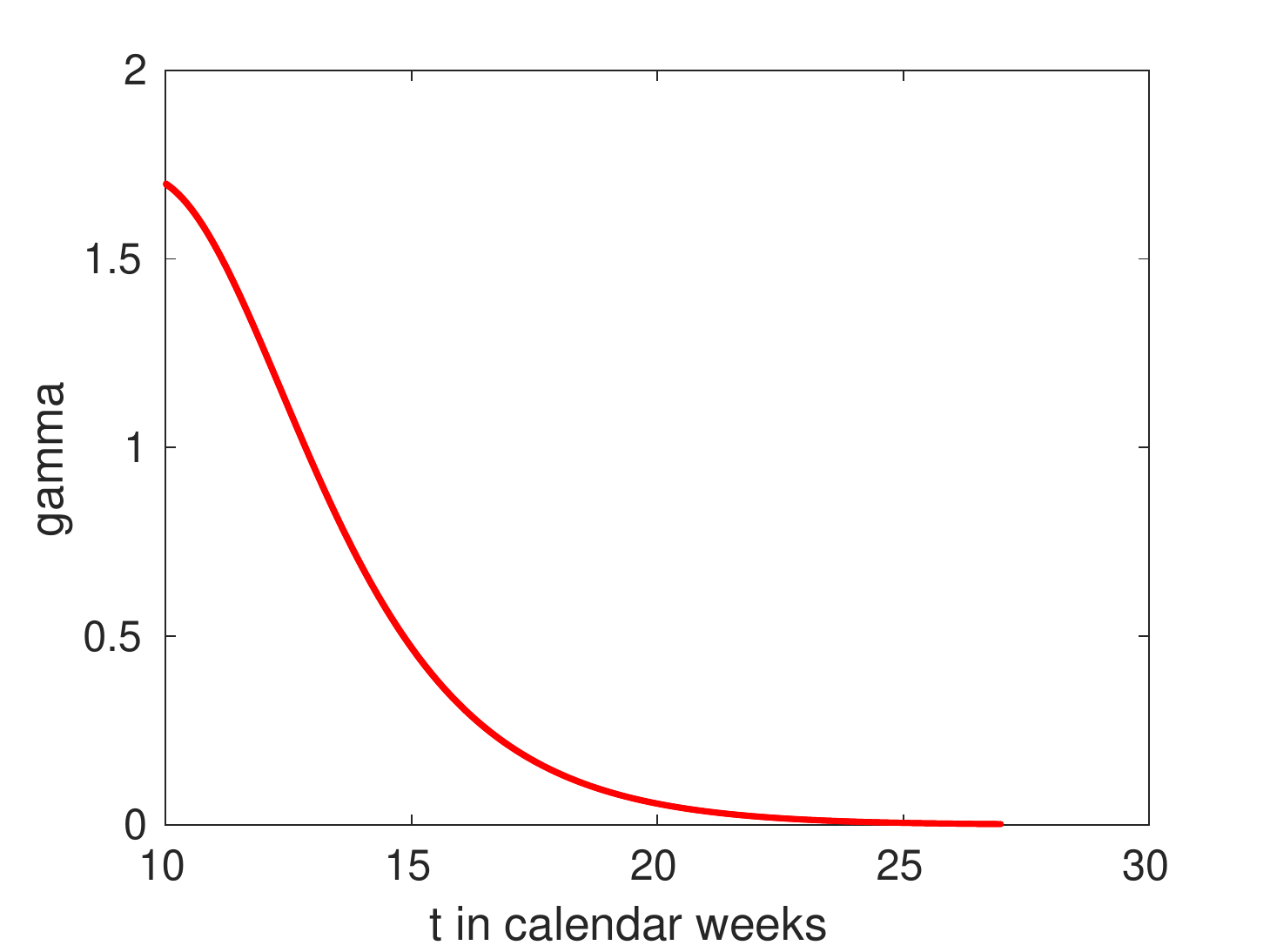} 
        \caption{ $\gamma(t)$} \label{fig:timing3}
    \end{subfigure}
     \hskip 1cm
     \begin{subfigure}[t]{0.35\textwidth }
    \centering
        \includegraphics[width=\linewidth]{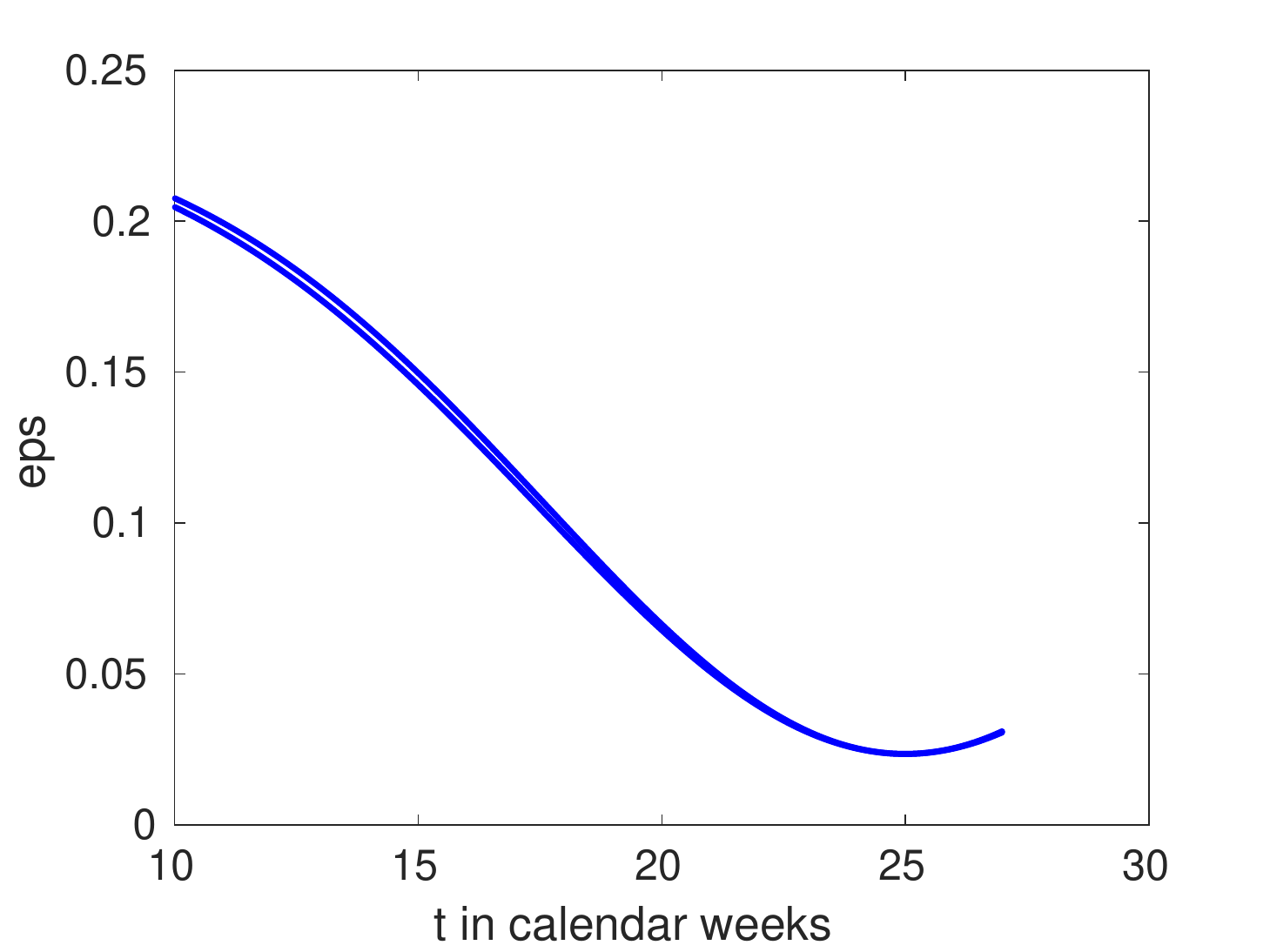} 
        \caption{Recovery rate $\varepsilon(t)$} \label{fig:timing3}
    \end{subfigure}
    \caption{Time dependence of $\tilde{I}(t)$, $R(t)$, $R_0(t)$, $\gamma(t)$ and $\varepsilon(t)$ within the eRG SIR model optimised to describe the French data for COVID-19. Solid lines are the model (shown as a 95\% confidence level band for the predictions)  and the dots with error-bars are the data.}
    \label{Francia}
\end{figure}

  The results for the French case can be found in Fig.~\ref{Francia} with the overall picture similar to the Italian one. The striking difference compared to the other countries is that the recovery rate decreases at late times. Given that the deaths in France amount to about 20\% of the total infected, such difference indicates that a more complete model (including the deaths compartment) is needed.    

\section{Conclusions}
  We generalised the epidemic Renormalisation Group framework to take into account the recovered cases and to be able to determine the time dependence of the reproduction number. At the same time we show that the eRG framework can be embedded into a SIR model with time-dependent coefficients. Interestingly the resulting infection rate $\gamma(t)$ is a smooth curve with a maximum at early times while rapidly plateauing at large times. This is a welcome behaviour since it encodes the slow down in the spreading of the disease at large times coming, for example, from social distancing. 
  
  We then move to confront the model to actual data by considering the spread of COVID-19 in the following countries: Denmark, Germany, Italy and France. We show that the overall approach works rather well in reproducing the data. Nevertheless the interpretation for the recovery rate is natural for Denmark and Germany while it requires to add the deaths compartment for Italy and France. The reason being that for the first two countries the number of deaths is below 5\% of the number of infected cases while it is above 15\% for France and Italy.  We therefore expect that this compartment will be relevant to include for countries with similar number of deaths.  The extension of the eRG to include also the death compartment is a natural next step.

\bibliography{bibAS}
\bibliographystyle{JHEP-2-2}

\end{document}